\documentstyle[sprocl]{article}

\def\be{\begin{equation}}
\def\ee{\end{equation}}
\def\bea{\begin{eqnarray}}
\def\eea{\end{eqnarray}}

\begin{document}


\title{GUT GENEALOGIES FOR SUSY SEESAW HIGGS \footnote{Talk presented at SUSY01, Dubna, Russia , June 2001}}

\author{C.S.AULAKH}

\address{Dept. of Physics, Panjab University, Chandigarh,INDIA,
 E-mail: aulakh@pu.ac.in}


\maketitle\abstracts{We discuss the embedding of  R-Parity preserving 
Minimal Left Right Supersymmetric models into Pati-Salam and SO(10) GUTs.}

 Super-Kamiokande observations strongly suggest $|m_{\nu_{\mu}}-m_{\nu_\tau}| \sim .05 eV$ 
and so, barring fine tunings,  we expect   $m_{\nu}< 10^{-1} eV $.
 Then if the neutrino is a Dirac particle (i.e $\nu_R$ is light) the relevant yukawa
 $y_{\nu_{\tau}} \leq 10^{-13}$. Else superheavy $\nu_R$  
deouple from the low energy dynamics leaving a light majorana $\nu_L$.
  Then the leading ($d=5$)  
  operator for neutrino mass ($(H^{\dagger}L)^2$) occurs with coefficient $M^{-1}$
with  $M \geq 10^{15} GeV$. Within the seesaw scenario\cite{seesaw,mohsen}  
$M_{\nu_R}\sim M  $.

New Physics at large scales motivates  Supersymmetry to maintain the structural stability of the 
theory against radiative disruption.
$M_{\nu_R}\sim M_{GUT}$  raises  hopes of a window into GUT physics via
 neutrino  masses and vice versa\cite{Patiwilbabu}.
 Spontaneous, renormalizable $M_{\nu_R}$ implies  
 a Higgs with $L=2$ is needed.  The $\nu_R$ 
field makes  $B-L$ anomaly free and  thus  gaugeable. The MSSM loses 
the     B,L symmetries of the SM  
due to the sfermions which allow B,L violating yukawas  which imply  catastrophic
nucleon decay. Imposing  matter parity ($P_M=(-)^{3(B-L)}$) or equivalently
R parity ($R_p=P_M (-)^{2S}$)\cite{rabirpar} is the nearly unique\cite{ibanross} cure. It restores 
B,L at least at the renormalizable level. Protection against gravitational 
violation then strongly motivates gauging of $B-L$. If only $B-L$ even fields
 get vevs $R_p$ is preserved. In fact one can prove\cite{aulben,rensee}that 
Supersymmetric theories with a renormalizable seesaw mechanism imply that $R_p$ will be
exact in the effective MSSM (with neutrino masses) after heavy $\nu_R$s
 are integrated out : susy-seesaw based on B-L even 
vevs leaves the low energy theory with an Lepton number symmetry violated only
 by the neutrino mass operator. So   sneutrino vevs lead to a unnaceptably
light ($<<M_Z/2$) would be Majoron which is emphatically ruled out by the Z width.

 Left-Right symmetric theories\cite{lrmodels} with 
gauge group  $G_{LR}\equiv SU(3)_c \times SU(2)_L \times SU(2)_R \times U(1)_{B-L}$ naturally gauge B-L\cite{rabimarsh}.
Parity is only spontaneously broken. The Higgs doublets of the MSSM are
in a $(1,2,2,0)$  while the $U(1)_{B-L}$
breaking is via  
$ \Delta = (1,3,1,2)  ,   \overline{\Delta}  = (1,3,1,-2), 
\Delta_c = (1,1,3,-2), \quad \overline{\Delta}_c = (1,1,3,2) $. These pairings ensure anomaly cancellation 
and also permit susy preserving breaking to the SM gauge group.

 Important variants contain a parity odd singlet(POS) field $\sigma$
which couples to the $\Delta$  fields as $W= \sigma (\Delta  \overline{\Delta} -  \Delta_c  \overline{\Delta}_c ) 
+ M \sigma^2 $. Its vev introduces spectrum asymmetry and  so also (via running of couplings ) coupling asymmetry.
The POS based LR Susy model initially introduced \cite{cvetic} 
broke charge\cite{kuchimoha}  for  generic  high $M_R$ . 
Minimal Left Right Symmetric models (MSLRMs) cure\cite{aulben} this  by 
introducing B-L neutral triplets $ \Omega(1,3,1,0) ,\Omega_c(1,1,3,0)$ or
using\cite{aulalej}.
 superpotentials $W=\Delta^4/M +....$.
 The $<\Omega^c>$   breaks $SU(2)_R \rightarrow U(1)_R$ followed by $U(1)_R\times U(1)_{B-L}\rightarrow U(1)_Y $
 breaking via the right handed $\Delta^c$s. This permits a 
heirarchy $M_R>> M_{B-L}>> M_{B-L}^2/M_R >> M_W$ , where the importance  of the geometric scale is
 that entire left handed 
triplets have mass at that scale. The non-renormalizable case is even more dramatic. 
$<\Delta^c>\sim \sqrt{m_{\Delta} M }\sim M_R$ leads to masses $\sim M_R^2/M$ for the fields 
$(\delta_c^0+\bar\delta_c^0),\delta_c^{++},\bar \delta_c^{++}, \Delta, {\overline \Delta}, H', {\overline H}'$.
Such light or intermediate mass scale multiplets are  generic\cite{surviv}  and  provide important constraints
 on LR models. {\it{The low energy effective theory of such models is 
the MSSM with exact R-parity and seesaw neutrino masses}} .
 
 How may we embed   MSLRMs  into SUSY GUTs so that they  may 
inherit  exact R-parity for the low energy theory even after $B-L$
is  broken and neutrino masses arise ?   An $SO(10)$ GUT based on the
Higgs  multiplets $45,54,126,{\overline {126}}$ which  has   MSSM with R-parity
 as its low energy limit
was given\cite{so10} but the relationship
between the MSLRMs introduced above (and their variants which include the POS) is quite rich..

MSLRMs embed easily in the Pati-Salam model with
 gauge group $SU(2)_L$\hfil\break $\times SU(2)_R\times SU(4)_C$. The fermions embed  into $(2,1,4) \oplus
(1,2,{\bar{4}})$ multiplets while the $\Delta$s with $B-L=\pm 2$  extend to SU(4) 10-plets.
 The breaking  $SU(4)\rightarrow SU(3) \times U(1)_{B-L}$ is via   $(1,1,15)$ . Since 
the LR  symmetry is external to the Gauge group the POS  is an additional
 singlet (i.e not in (1,1,15) as in SO(10))
(see below). The  basic MSLRM's (no POS) embed using
$SU(4)$ singlet $\Omega$s or else allowing
$W\sim \Delta^4/M$ . As before one gets intermediate mass multiplets in both cases .
POS variants are also easy to write down.

  Embedding MSLRMs in the SO(10) GUTs is more subtle.   PS  $(2,1,4) \oplus
(1,2,{\bar 4})$ matter fermions of each generation 
embed  into $16$ plets of SO(10), $\Delta, \Delta^c$ pairs embed into $126,{\overline {126}}$ .
A renormalizable superpotential that breaks SO(10) to the Pati-Salam group and then to   the LR symmetric group 
(Chain a)) or $G_{214}$ (Chain b)) and finally to the MSSM via 126 vevs 
was   constructed\cite{so10}. 
  Chain a) proceeds via  $<(1,1,15)>\quad \subset 45$ and Chain b)  via $<(1,3,1)>$.
{LR} symmetry embeds in SO(10) as $D_{LR}\sim  \sigma_{23} \sigma_{67}$. Then  
 $<(1,1,1,0)> \quad \subset \qquad <(1,1,15)>$ is Parity {\bf{odd }} and  so the  theory below 
the PS breaking scale is {\it not} left-right symmetric though $G_{LR}$ is . The light
Higgs below  $M_{PS}$ descend from  decuplets or  (1,1,15) and  include no $\Omega$s but have 
POS (1,1,1,0) and a colored companion (1,1,8,0).  This case thus has 
non-renormalizable terms (induced by integrating out the heavy superfields) 
and a POS  i.e it  embeds
  the POS variant of the second MSLRM. One loop RG anaysis of   gauge 
  running   shows $M_{PS}, M_R, M_{B-L}\sim 10^{-1} M_X$  which  
 is {\it{lowered}} with them  and hence the 
  $d=6$ contribution to nucleon decay re-emerges even in the SUSY case.
  Typically $M_R, M_C, M_{B-L} \sim 10^{13.5} - 10^{14.5} GeV$
  while $M_X\geq 10^{15.5} GeV$ (from requirements of nucleon stability).

Chain b) utilizes the $\Omega^c (1,3,1)\in 45$ to break
   $SU(2)_R\rightarrow U(1)_R$.
  It   embeds   the PS version of the model with $\Omega$s
into SO(10). Thus the POS and $\Omega$ variants are mutually exclusive in GUTs based on the 45 of
SO(10).

To decouple parity and $SU(4)$ breaking needs SO(10) models 
  based upon\cite{aulmohso10}   210 
  which has a parity odd (1,1,1)   submultiplet and a  Parity {\it even} (1,1,1,0)   multiplet.
 The parity even $(1,1,1,0)\subset (1,1,15)\quad \subset 210$  will break to  $G_{LR}$ group
maintaining $D_{LR}$ symmetry while the parity odd $<(1,1,1)>$  violates
$D_{LR}$.  

 As in  MSLRM's  with higher dimensional terms  , one may also construct models
based on the 45 and 126 alone (i.e no 54) using $d>4$ terms, details will appear.
  
To conclude  :   MSLRMs  with  exact $R_P$ embed   non trivially into various SO(10) and PS GUTs  if
the POS models  are kept in mind.
 The one loop RG analysis of 
the gauge couplings shows that the the intermediate scales lie close
to the unification scale so that a kind of ``SU(5)
 conspiracy''  holds. This is due to the fact that violations of the survival principle in SUSY GUTs keep certain 
colored and charged  multiplets lighter than one would expect and their contributions significantly constrain the
 possibility of intermediate scales. 
 
{\bf{Acknowledgments}}
This talk is largely a report of work done in collaboration with 
 B.Bajc, A.Melfo, A.Rasin and G.Senjanovic.


\centerline{\bf{References}}
\begin{thebibliography}{99}

\bibitem{seesaw} M.~Gell-Mann, P.~Ramond and R.~Slansky, 
in {\it Supergravity}, eds. P.~van~Niewenhuizen and D.Z.~
Freedman (North Holland 1979); T.~Yanagida, in Proceedings 
of {\it Workshop on Unified Theory and Baryon number in the 
Universe}, eds. O.~Sawada and A. Sugamoto (KEK 1979); 
R.N.~Mohapatra and G.~Senjanovi{\'c}, Phys. Rev. Lett. 
{\bf 44}, 912 (1980).
\bibitem{mohsen} 
R.N.~Mohapatra and G.~Senjanovi\'c, Phys. Rev. {\bf D23},165 (1981); 
M.~Magg and Ch.~Wetterich, Phys. Lett. {\bf B94} 61, (1980).


\bibitem{Patiwilbabu}K.S. Babu   , Jogesh C. Pati  , Frank Wilczek, Nucl.Phys. B566 (2000) 33-91.
\bibitem{rabirpar} R.N. Mohapatra, Phys. rev. {\bf{D34}},
3457(1986); A.Font, L. Ibanez and F. Quevedo, Phys. Lett.
{\bf{B228}}, 79 (1989).

\bibitem{ibanross} L.E,Ibanez and G.G.Ross, Nucl. Phys. {\bf{B368}}(1992)3.
\bibitem{lrmodels} J.C. Pati and A. Salam, Phys. Rev.
{\bf{D10}},275 (1974);\hfil\break
R.N. Mohapatra and J.C. Pati, Phys. Rev. {\bf{D11}}, 566, 2558
(1975);\hfil\break 
G.Senjanovic and R.N. Mohapatra, Phys. Rev. {\bf{D12}}, 1502
(1975). 

\bibitem{cvetic} M.~Cveti\v{c},
  Phys. Lett. {\bf 164B}, 55 (1985).
\bibitem{kuchimoha} R.~Kuchimanchi and R.~N. Mohapatra,
  Phys. Rev. {\bf D48}, 4352 (1993), hep-ph/9306290,
  Phys. Rev. Lett. {\bf 75}, 3989 (1995), hep-ph/9509256.
\bibitem{aulben}Charanjit S.~ Aulakh, Karim Benakli, Goran Senjanovic,\hfil\break
 Phys.Rev.Lett.{\bf{79}}, 2188, 1997,  hep-ph 9703434.

 \bibitem{aulalej}C.S.~Aulakh, A.~Melfo, A.~Ra\v{s}in 
and G.~Senjanovi\'c, Phys. Rev {\bf D58}, 115007 (1998), 
hep-ph/9712551.
  \bibitem{rensee} C.S.~Aulakh, A.~Melfo, A.~Ra\v{s}in and 
G.~Senjanovi\'c, Phys. Lett. {\bf B459}, 557 (1999).

\bibitem{surviv}  C.S.~Aulakh, B.~Bajc, A.~Melfo, A.~Ra\v{s}in and
G.~Senjanovi\'c, Phys. Lett. {\bf B460}, 325 (1999), hep-ph/9904352.
 
\bibitem{so10}C.S.~Aulakh, B.~Bajc, A.~Melfo, A.~Ra\v{s}in and
G.~Senjanovi\'c, Nucl. Phys. {\bf{B597}}, 89 (2001), hep-ph/0004031. 
 \bibitem{aulmohso10}  C.S. Aulakh, R.N. Mohapatra, Phys.Rev.D28:217,1983.
\end{thebibliography}
\end{document}